\documentclass[a4paper,11pt]{article}
\usepackage{pos}
\usepackage{mhchem}

\graphicspath{{immagini/}} 

\title{Searching for an eco-friendly gas mixture for the ALICE Resistive Plate Chambers}

\author*[j]{Luca Quaglia on behalf of the ALICE and ECOgas@GIF++ collaborations}
\author[a]{R. Cardarelli, B. Liberti, E. Pastori, G. Proto}
\author[b]{G. Aielli, P. Camarri, A. Di Ciacco, L. Di Stante, R. Santonico}
\author[c]{G. Alberghi, D. Boscherini, A. Bruni, L. Massa, A. Polini, M. Romano}
\author[d]{L. Benussi, S. Bianco, L. Passamonti, D. Piccolo, D. Pierluigi, A. Russo}
\author[e]{M. Ferrini, G. Saviano}
\author[f]{M. Abbrescia, L. Congedo, M. De Serio, G. Galati, G. Pugliese, S. Simone}
\author[g]{P. Salvini}
\author[h]{A. Samalan, M. Tytgat, N. Zaganidis}
\author[i]{J. Eysermans}
\author[j]{A. Ferretti, M. Gagliardi, L. Terlizzi, E. Vercellin}
\author[k]{P. Dupieux, B. Joly, S.P. Manen}
\author[l]{M. Corbetta, R. Guida, B. Mandelli, G. Rigoletti}
\author[m]{M. Barroso}
\author[n]{M.C. Arena, M. Verzeroli}
\author[o]{A. Pastore}
\author[o,p]{R. Aly}

\affiliation[a]{INFN, Tor Vergata, Rome, Italy}

\affiliation[b]{Dipartimento di Fisica di Roma Tor Vergata, Rome, Italy}

\affiliation[c]{Sezione INFN di Bologna, Bologna, Italy}

\affiliation[d]{Laboratori Nazionali di Frascati dell'INFN, Frascati, Italy}

\affiliation[e]{Dipartimento di Ingegneria Chimica Materiali Ambiente, Sapienza Università di Roma, Rome, Italy}

\affiliation[f]{Dipartimento di Fisica di Bari e sezione INFN di Bari, Bari, Italy}

\affiliation[g]{Sezione INFN di Pavia, Pavia, Italy}

\affiliation[h]{Dept. of Physics and Astronomy, Ghent University, Proeftuinstraat 86 B-9000 Ghent, Belgium } 

\affiliation[i]{MIT, Cambridge, Massachussets, U.S.A.}

\affiliation[j]{Università di Torino and INFN, sezione di Torino via Giuria 1, 10125 Torino, Italy}

\affiliation[k]{Clermont Univeristé, Université Blaise Pascal, CNRS/IN2P3, Laboratoire de Physique Corpusculaire BP 10448, F-63000 Clermont-Ferrand, France}

\affiliation[l]{CERN, Geneve, Switzerland}

\affiliation[m]{Universidade do Estdo do Rio de Janeiro, Rio de Janeiro, Brazil}

\affiliation[n]{Università degli Studi di Pavia, Pavia, Italy}

\affiliation[o]{Sezione INFN di Bari, Bari, Italy}

\affiliation[p]{Helwan University, Ain Helwan 11795 Cairo, Egypt}

\emailAdd{luca.quaglia@cern.ch}

\abstract{
	\textit{ABSTRACT}: Resistive Plate Chambers (RPCs) are gaseous detectors with parallel plate geometry and resistive electrodes, widely employed at the LHC. In ALICE (A Large Hadron Collider Experiment) 72 RPCs are installed in the forward muon spectrometer and provide muon identification. 
	
	The ALICE RPCs are operated with a mixture of 89.7\% $C_{2}H_{2}F_{4}$, 10\% i-$C_{4}H_{10}$ and 0.3\% $SF_{6}$. $C_{2}H_{2}F_{4}$ and $SF_{6}$ are fluorinated greenhouse gases (F-gases) with a high Global Warming Potential (GWP). New European Union regulations have imposed a progressive phase-down of the production and usage of F-gases, aiming to cut down their emission by two thirds in 2030 with respect to 2014.\\
	Even though research activities are excluded from these regulations, the F-gases phase-down will inevitably increase their price and CERN is also aiming to cut down on its emissions. For these reasons it is crucial to find a more eco-friendly gas mixture for RPCs by the time of the LHC long shutdown 3, foreseen in 2026. Since $C_{2}H_{2}F_{4}$ is the main contributor to the mixture’s GWP an extensive R\&D process has started to replace it with tetrafluoropropene ($C_{3}H_{2}F_{4}$), due to its chemical similarity with $C_{2}H_{2}F_{4}$ and its low GWP (around 7). Preliminary tests with cosmic rays have shown promising results in terms of detector performance. The next step is to study the long-term behavior of RPCs operated with these new gas mixtures (aging studies). Since this is a subject of interest for all (and not only) the LHC experiments, a collaboration, ECOgas@GIF++, was setup to carry out joint studies. 
	
	Among others, small-size, ALICE-like RPCs were installed at the Gamma Irradiation Facility at CERN, where they are exposed to a strong radiation field, coming from a 12.5 TBq $^{137}$Cs source, which allows one to simulate many years of operation in a relatively short time. The facility also provides a muon beam at specific times of the year, which can be used to study the detector performance (e.g. efficiency and cluster size) during and after irradiation.
	
	This poster reports the current status of the measurements, focusing on the preliminary results of the irradiation campaign and of beam tests carried out in 2021.}

\FullConference{%
  The Tenth Annual Conference on Large Hadron Collider Physics - LHCP2022\\
  16-20 May 2022\\
  online
}

\begin{document}
\maketitle

\section{Introduction}
Resistive Plate Chambers (RPCs) are gaseous particle detectors widely employed in the muon systems of high energy physics experiments. In particular, thanks to their timing resolution of 1-2 ns and spatial resolution in the order of few mm, they can be employed for triggering purposes. Furthermore, due to their comparatively low cost, they are suitable to cover large areas. This is the case for the ALICE \cite{alice} (A Large Ion Collider Experiment) detector at the CERN Large Hadron Collider (LHC), where a total of 72 RPCs are employed for muon identification and cover an area of around 140 m$^{2}$. These detectors are operated with a mixture of  89.7\% \ce{C_{2}H_{2}F_{4}} (also known as R134a), 10\% \ce{i-C_{4}H_{10}} and 0.3\% \ce{SF_{6}}. Such a mixture  provides a working voltage around 9.8 kV and excellent performances \cite{mtr_perf}.

Out of the three gases in the mixture, \ce{C_{2}H_{2}F_{4}} and \ce{SF_{6}} are greenhouse gases, meaning that they trap great amounts of heat in the atmosphere for an extended period of time. New European Union regulations \cite{regulations} have imposed a progressive phase down in the production and use of such gases, reducing their availability and increasing their price. Even though scientific laboratories are not explicitly included in these regulations, CERN has asked its scientific community to start an effort to find a new, more eco-friendly, gas mixture to be used in RPC detectors.  

In order to quantify how harmful a gas is for the environment, the global warming potential (GWP) has been introduced. This is a number that compares the amount of heat trapped in the atmosphere by a ton of certain gas (over a given period of time, usually 100 years) if compared to a ton of \ce{CO_{2}}. By definition GWP(\ce{CO_{2}}) = 1. For reference GWP(\ce{C_{2}H_{2}F_{4}}) $\simeq$ 1400 and GWP(\ce{SF_{6}}) $\simeq$ 22800.

Since each of the employed gases has a unique role in signal development, it is not an easy task to find a suitable replacement. For this reason, most of the efforts have been focused on the replacement of \ce{C_{2}H_{2}F_{4}}, the main contributor to the gas mixture GWP. A possible candidate for the replacement has been identified in Tetrafluoropopene \cite{hfo} (\ce{C_{3}H_{2}F_{4}}), also known as HFO-1234ze (HFO in the following). It has a very similar chemical structure to R134a, with a much lower GWP of around 7. Furthermore, this gas is being employed as an industrial replacement for R134a, making it readily available and cheap.

HFO has a smaller effective first Townsend coefficient, making it impossible to simply replace the R134a \textit{in toto}, since the working voltage would move to over 15 kV. Such voltages cannot be tolerated by the existing high voltage infrastructures and detectors. Hence, many preliminary tests with cosmics, have been performed in order to find possible ways to work with this gas. In particular, it has been found that diluting HFO with \ce{CO_{2}} gives the best performance in terms of operating voltage and efficiency \cite{co2}. A few promising gas mixtures have been identified and the next step is to study the detector's response in more controlled environments (i.e. beam tests), and, at the same time, their long-term behavior (i.e. aging studies) when operated with these new gas mixtures.

\section{Experimental setup}

A cross-experiments collaboration (ALICE, ATLAS, CMS, CERN EP-DT and SHiP/LHCb) was established to perform the tests described above, joining forces to reach the common goal of finding a new gas mixture for RPC detectors. Each group provided an RPC prototype, all installed in a common experimental setup at the CERN Gamma Irradiation Facility (GIF++) \cite{gifpp}. 

This facility is equipped with a 12.5 TBq \ce{^{137}Cs} source, which provides a uniform gamma irradiation field that can be used to simulate a high background environment to speed the aging process, allowing one to simulate several years of irradiation in a reduced amount of time. In addition, during dedicated beam periods, a muon beam of 100 GeV/c is produced by a hadron beam (circulating in the CERN SPS) impinging on a target. The peculiarity of the GIF++ is the possibility to combine the muon beam with the Cs source, enabling the study of the detector performance under various background conditions, hence giving insights into the RPC rate capability.

The objective of this contribution is on the detector that was provided by the ALICE collaboration. This is a small (50x50 cm$^{2}$) ALICE-like RPC, with 2 mm thick bakelite electrodes and 2 mm thick (single) gas gap. Signal amplification and discrimination were provided by the FEERIC \cite{feeric} front-end boards, currently employed in ALICE. The readout was executed through a CAEN TDC (model V1190a) and the trigger was provided by the coincidence of 4 scintillators coupled with photomultipliers  (two placed outside the irradiation facility and two inside) for a total trigger area of 10x10 cm$^{2}$.

A total of three gas mixtures were tested, the CMS/ATLAS standard gas mixture (STD)\footnote{Note that ALICE standard gas mixture could not be used for flammability reasons} (95.2\% \ce{C_{2}H_{2}F_{4}}, 4.5\% \ce{i-C_{4}H_{10}} and 0.3\% \ce{SF_{6}}) and two eco-friendly gas mixtures, ECO2 (60 \% \ce{CO_{2}}, 35 \% \ce{HFO}, 4 \% \ce{i-c_{4}H_{10}} and 1 \% \ce{SF_{6}}) and ECO3 (69 \% \ce{CO_{2}}, 25 \% \ce{HFO}, 5 \% \ce{i-c_{4}H_{10}} and 1 \% \ce{SF_{6}}) in the following, both with a GWP $\simeq$ 230.

\section{Results}

This contribution shows the latest results from a beam test carried out in October 2021. In particular, the detector was flushed with the standard gas mixture in order to have a reference measurement, which was then compared to the two ecogases. The Cs source can be shielded by means of a set of lead filters, allowing one to test the detectors at different background rates. For the upcoming LHC physics runs, the expected hit rate on the ALICE RPCs is around 100 Hz/cm$^{2}$, which includes a safety factor of 2 \cite{alicerun3}. 

\begin{figure}[h]
	\centering
	\begin{minipage}[b]{0.49\textwidth}
		\includegraphics[width=\textwidth]{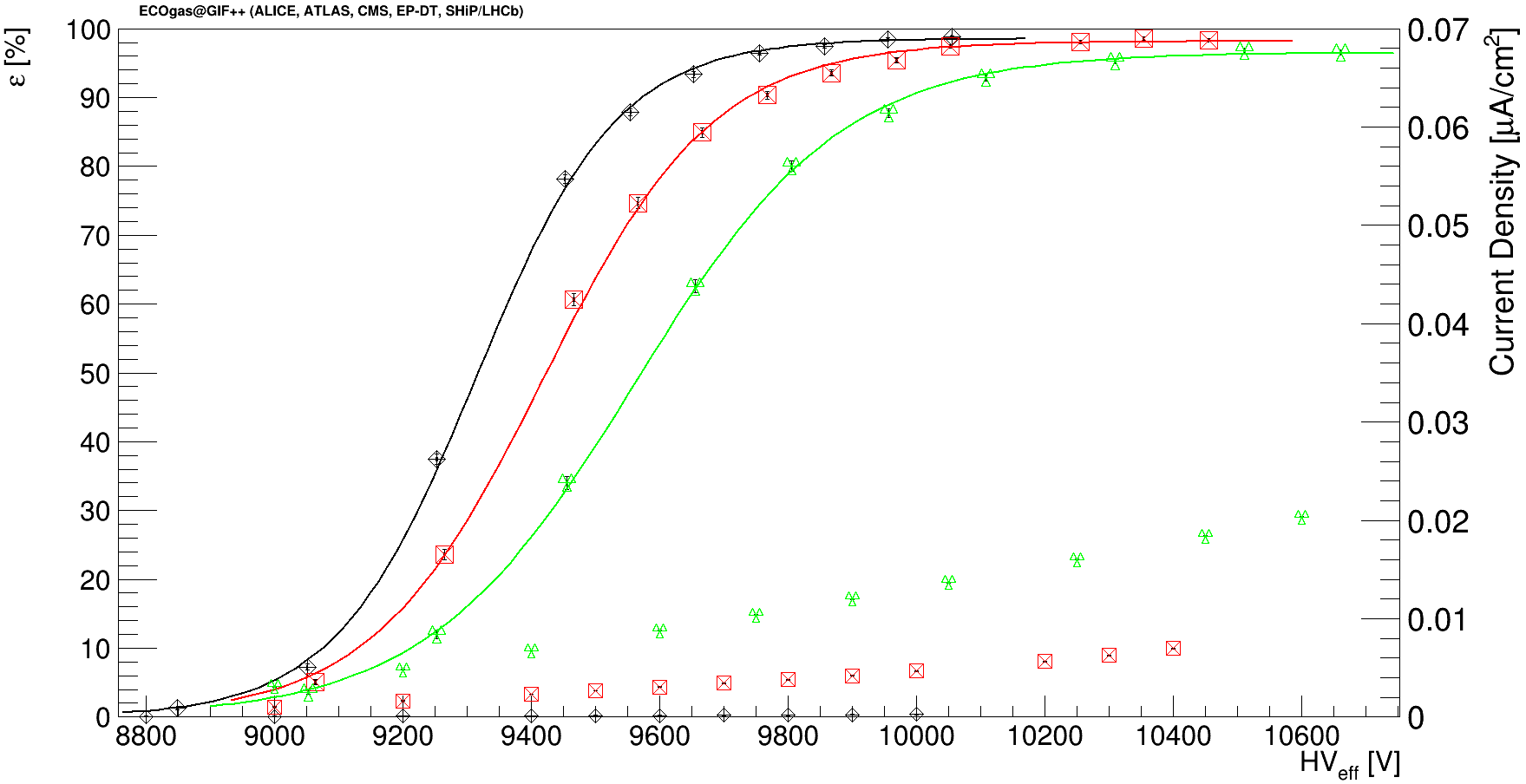}
		\caption{STD gas mixture}
		\label{ALICE_std}
	\end{minipage}
\end{figure}

In figure \ref{ALICE_std}, the performance in terms of efficiency and absorbed current is shown for the standard gas mixture. Since the RPC was read out on both planes, with perpendicular strips, it was considered to be efficient if, for a given trigger, at least one strip per plane had given a signal above the set threshold. Also, the absorbed current was measured and, in the figure, the current density (in $\mu$A/cm$^{2}$) is reported. The black line represents the measurement taken without irradiation (source OFF in the following), while the red and green ones show respectively the behavior of the RPC with a gamma hit rate of 100 and 300 Hz/cm$^{2}$.

It is possible to observe that for higher rates, the efficiency curves are shifted to higher voltages, due to a high drop in the high voltage caused by the higher rate of impinging photons. At the same time, the maximum value of efficiency is slightly reduced. These observations can be explained by taking into consideration the RPC rate capability.

\begin{figure}[h]
	\centering
	\begin{minipage}[b]{0.49\textwidth}
		\includegraphics[width=\textwidth]{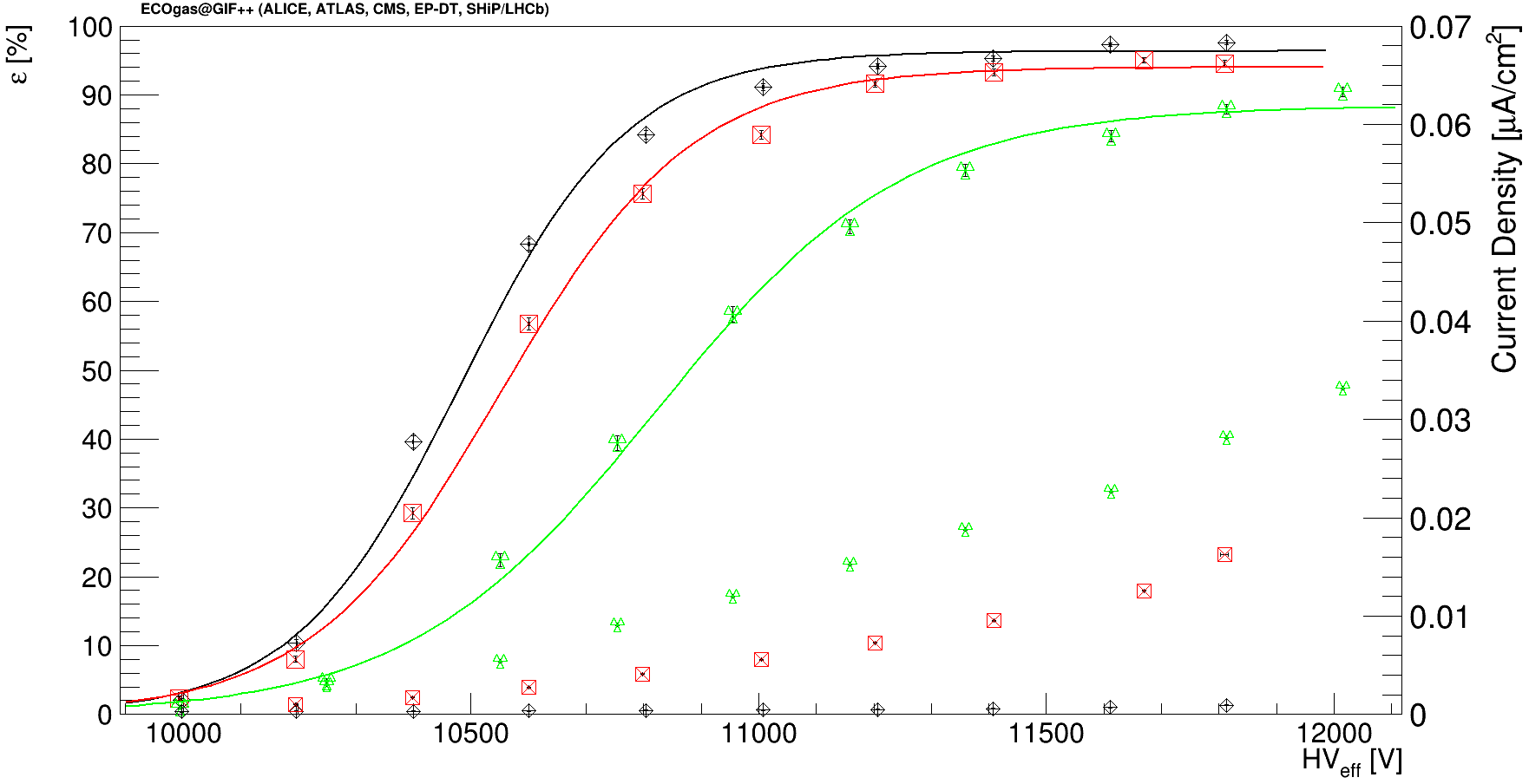}
		\caption{ECO2 gas mixture}
		\label{ALICE_eco2}
	\end{minipage}
	\begin{minipage}[b]{0.49\textwidth}
		\includegraphics[width=\textwidth]{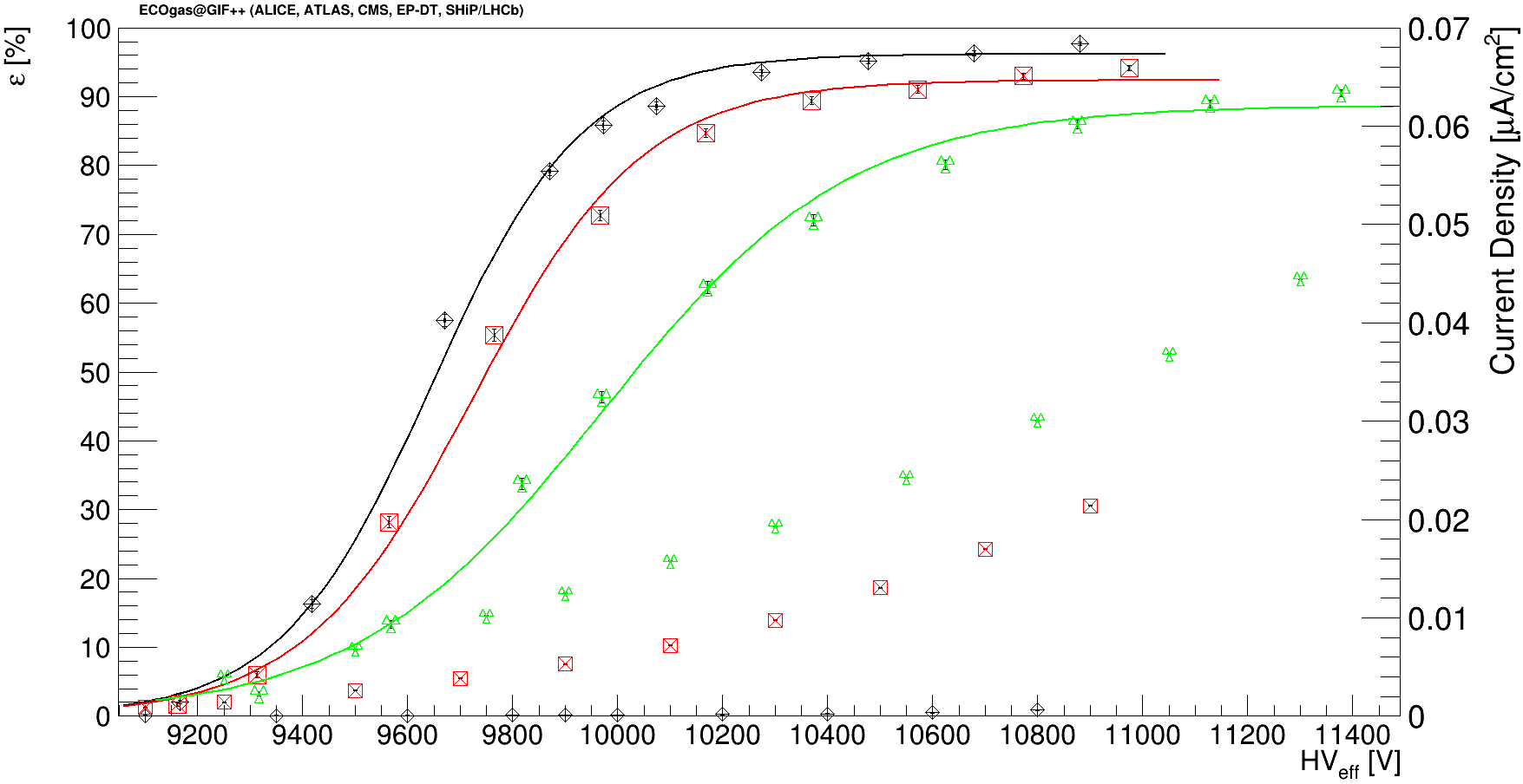}
		\caption{ECO3 gas mixture}
		\label{ALICE_eco3}
	\end{minipage}
\end{figure}

In figures \ref{ALICE_eco2} and \ref{ALICE_eco3} the results obtained under the same experimental conditions as discussed earlier for the standard gas mixture, are reported for the two ecogases under study.

The results with both ecogases are promising and a few observations can be made: the first one is that the source OFF efficiency curve is shifted towards higher voltages, concerning the standard gas mixture, due to the presence of HFO. In particular, for the ECO2 mixture the shift is 1.2 kV while for ECO3 is 0.4 kV. 

The effects of the RPC rate capability is even more visible concerning the standard gas mixture, yielding a greater shift and higher efficiency drop. However the situation at a rate of 100 Hz/cm$^{2}$ (red curve) is still acceptable. 

The last point that could be raised regards the current absorbed by the detector, in particular, it is doubled and this needs to be closely studied to investigate if it could lead to a deterioration effect on the detector performance.

\section{Conclusion}

A search for new eco-friendly gas mixtures for RPCs is ongoing. In this context, the ECOgas@GIF++ collaboration was established and systematic aging and beam test studies are ongoing. During a dedicated beam test period the performance of an ALICE-like prototype were tested with two eco-friendly gas mixtures and compared to a well-known \ce{C_{2}H_{2}F_{4}}-based gas mixture. The detector has shown satisfactory performance for the expected rate in ALICE during the upcoming LHC physics runs but the working point is moved to higher voltages and the absorbed current is doubled with respect to the reference gas mixture. For this reason, it is mandatory to carry out in-depth aging studies to better understand whether or not these issues have any long-term impact on the detector performances. 

\newpage

\bibliographystyle{ieeetr} 
\bibliography{refs} 

\end{document}